\documentclass[final,1p,times]{elsarticle}

\usepackage{graphicx}
\usepackage{color}

\usepackage{amssymb}
\usepackage{amsmath} %
\biboptions{sort&compress}
\usepackage{lscape}

\journal{Nuclear Instruments and Methods A}

\begin{document}
\begin{frontmatter}



\title{Measurement of Radon Concentration in Super-Kamiokande's Buffer Gas}

\author[A]{Y.~Nakano\corref{cor1}}
\ead{ynakano@phys.sci.kobe-u.ac.jp}
\author[B,C]{H.~Sekiya}
\author[B]{S.~Tasaka}
\author[A,C]{Y.~Takeuchi}
\author[D,C]{R.~A.~Wendell}
\author[E]{M.~Matsubara}
\author[B,C]{M.~Nakahata} 

\address[A]{Department of Physics, Kobe University, Kobe, Hyogo 657-8501, Japan}
\address[B]{Kamioka Observatory, Institute for Cosmic Ray Research, The University of Tokyo, Gifu 506-1205, Japan}
\address[C]{Kavli Institute for the Physics and Mathematics of the Universe (WPI), The University of Tokyo Institutes for Advanced Study, University of Tokyo, Kashiwa, Chiba 277-8583, Japan}
\address[D]{Department of Physics, Kyoto University, Kyoto, Kyoto 606-8502, Japan}
\address[E]{Information and Multimedia Center, Gifu University, Gifu 501-1193, Japan}

\cortext[cor1]{Corresponding author. Present address: Department of Physics, Kobe University, Kobe, Hyogo 657-8501, Japan. Tel: +81 78 803 5655; Fax: +81 78 803 5662.}

\begin{abstract}
To precisely measure radon concentrations in purified air supplied to the Super-Kamiokande detector as a buffer gas, 
we have developed a highly sensitive radon detector with an intrinsic background as low as $0.33\pm0.07$ $\mathrm{mBq/m^{3}}$.
In this article, we discuss the construction and calibration of this detector as well as results of its application to the measurement and monitoring of the buffer gas layer above Super-Kamiokande. 
In March 2013, the chilled activated charcoal system used to remove radon in the input buffer gas was upgraded.
After this improvement, a dramatic reduction in the radon concentration of the supply gas down to $0.08\pm0.07$ $\mathrm{mBq/m^{3}}$.
Additionally, the Rn concentration of the in-situ buffer gas has been measured $28.8\pm1.7$ $\mathrm{mBq/m^{3}}$ using the new radon detector. Based on these measurements we have determined that the dominant source of Rn in the buffer gas arises from contamination from the Super-Kamiokande tank itself.
\end{abstract}

\end{frontmatter}


\section{Introduction}

Super-Kamiokande (Super-K) is a water Cherenkov detector containing 50,000 t of  ultra-pure water and 11,129 20-in-diameter photomultiplier tubes that observe the light produced by charged particles traversing the detector's inner volume \cite{paper_SK, SK_calib}. 
To study the interactions of neutrinos produced in the atmosphere, solar interior, and cosmos,  the detector is situated 1,000 m underground (2,700 m water equivalent) in Gifu prefecture, Japan.
 With the cosmic ray background largely suppressed at this depth, Super-K is able to observe the recoil electron produced by $\mathrm{^{8}B}$ solar neutrinos $(E_{\nu}\le15$ MeV) scattering elastically with electrons in the detectors water.
 As these recoiling electrons preserve the direction of their parent neutrinos, their measurements played a critical role in establishing neutrino oscillations as the solution to the ``solar neutrino problem'' \cite{bahcall_1973}.
 While our understanding of solar neutrinos has improved dramatically since then~\cite{sk1_solar, sno}, the predicted distortion of the energy spectrum induced by the effects of solar matter on their oscillation, the so-called Mikheyev-Smirnov-Wolfenstein (MSW) mechanism~\cite{msw1, msw2}, is yet to be confirmed completely. Although current data is well described by the predictions of the MSW model at low energies, where so-called vacuum oscillations are dominant~\cite{home,sage,gno,bore1,bore2,bore3,bore4,kamland1} as well as high energies, where matter effects dominate~\cite{sk1,sk2,sk3,sk4_solar,sno1,sno2,sno3,sno4,kamland2}, the shape of the electron neutrino survival probability in the transition region between these two energy regimes has not been determined yet. 
At present several alternative models have been proposed, such as sterile neutrino models~\cite{sterile1,sterile2}, mass varying neutrinos~\cite{massvary}, and non-standard interactions~\cite{non1,non2}, all of which explain the current experimental data but provide different predictions in the transition region.
In order to different between these and the standard MSW effect, precision solar neutrino measurements at intermediate energies are needed.

At the same time, Super-K's underground location makes it subject to similar-energy backgrounds produced by radioactive elements in the host mine's air, water supply, and surrounding rock. 
 One of major backgrounds in the kinetic energy region $E\le6$ MeV is electron events produced by radon ($\mathrm{^{222}Rn}$).
 These backgrounds are partially mitigated by continuously circulating and purifying the detector water and by supplying a layer of purified air between the top of the steel structure defining the detector and the tank water as a buffer gas \cite{paper_SK}.
 Although recent upgrades to both of these systems~\cite{SK_calib} and the Super-K front end electronics~\cite{qbee} have resulted in an analysis energy threshold of $3.5$ MeV (electron kinetic energy)~\cite{sk4_solar}, in order to observe the MSW spectrum distortion, the threshold must be reduced further.
 Currently, the threshold is limited by a few $\mathrm{mBq/m^{3}}$ of Rn estimated 
 to exist in the $4.0$--$6.0$ MeV energy region of the analysis~\cite{Rn_sk, nakano}.
 Reducing the threshold requires that this concentration be reduced to less than $0.1$ $\mathrm{mBq/m^{3}}$ and motivates careful analysis of potential radon sources.
 Dissolution of Rn into the detector water from the buffer gas layer is one such source.
 In this article, we discuss studies of the Rn concentration in the detector buffer gas layer using a new detector designed for precise measurements. 

This paper is organized as follows.
 In Section \ref{sec2}, the designs of the new 80 L Rn detector and its data acquisition system are described.
 In addition, the detection method and typical pulse height spectra are discussed.
 In Section \ref{sec3}, the developed calibration setup and the calibration results are described.
 In Section \ref{sec4}, the detector's intrinsic background level is briefly discussed.
 In Section \ref{sec5}, a Rn monitoring system to measure the Rn concentration in the buffer gas as well as an improved Rn-reduced air system in Super-K are described.
 Then, the results of measurements taken from December 1$^{\mathrm{st}}$, 2012 to April 9$\mathrm{^{th}}$, 2014 are presented.
 Finally, the study is concluded and its future prospects are discussed in Section \ref{sec6}.

\section{ Ultra sensitive 80 L Rn detector}
\label{sec2}
While previous Rn measurements at Super-K used detectors of size 70~L~\cite{Radon_takeuchi_1999} and 700~L~\cite{Radon_takeuchi_2003}, limitations in their measurement capabilities led to the development of an 80~L detector in the current study \cite{PTEP_hoso}.
 Like its predecessors, the Rn concentration of the sample air filling the detector volume is measured by electrostatically trapping $\mathrm{^{222}Rn}$ decay products, namely, $\mathrm{^{210}Po}$ (5.30 MeV), $\mathrm{^{214}Po}$ (7.69 MeV), and $\mathrm{^{218}Po}$ (6.00 MeV), and then, a PIN photodiode is used to observe their subsequent  $\alpha$ decays.
 With both the highest $\alpha$ energy of the three daughters and a high efficiency for electrostatic collection, $\mathrm{^{214}Po}$ is used as the primary signal for this study.
 The sample's Rn concentration is proportional to the number of observed $\alpha$ decays from this daughter.

\subsection{Structure of the new radon detector}

\begin{figure}[h]
\centering
\includegraphics[width = 80mm]{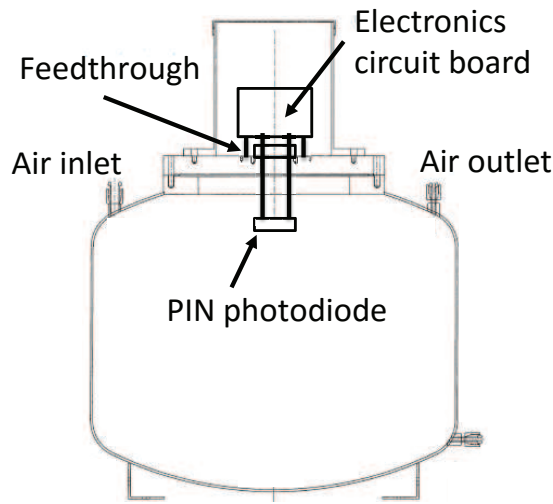}
\caption{The structure of the 80 L Rn detector.}
\label{rn_det}
\end{figure}

The detector consists of an 80~L stainless steel vacuum vessel with a PIN photodiode connected to a ceramic feedthrough mounted on the top flange of the vessel. The photodiode is offset from the bottom of the feedthrough by 5~cm. A high-voltage divider and amplifier circuitry rests atop this flange and is connected to the external end of the feedthrough. In this configuration, the p-layer of the PIN photodiode is biased at $-1.9$ kV (discussed later in Section 3.3), while the vessel is held at ground. Fig. \ref{rn_det} shows the structure of the detector.

Measurement limitations suffered by predecessor detectors due to materials used in their construction have been largely overcome in designing the new detector.
For instance, Viton$^{\textcircled{\footnotesize{R}}} $ O-rings, which were used as gaskets, have been replaced in the current design with metal gaskets because Viton$^{\textcircled{\footnotesize{R}}} $ is known to emanate Rn.
Similarly, acrylic flanges used in earlier designs have been replaced with knife-edge flanges with metal gaskets (CF flanges) to reduce the lower limit of the vessel's vacuum from 10~kPa to $<1.0\times10^{-5}$~Pa.
To further reduce backgrounds intrinsic to the detector itself, the inner surface of the stainless vessel has been electropolished to a roughness of less than $0.8$ $\mu$m. It should be noted that while it is advantageous to improve the performance the present detector's 700~L predecessor, electropolishing its larger surface area 
and fitting it with a CF flange with a 100~cm are both cost and labor intensive. 
For this reason, the present work seeks to establish the capabilities of these upgrades with the smaller 80~L detector.

In the current design, the PIN photodiode has been upgraded to a HAMAMATSU-S3204-09 with dimensions 18 mm $\times$ 18 mm, whose window has been removed to reduce energy loss as $\alpha$ particles strike the photodiode.
Rn emanation from the high-voltage feedthrough has additionally been reduced by switching its ceramic to high-purity alumina ($\mathrm{Al_{2}O_{3}}$).

\subsection{Electronics circuit and Data acquisition system}
The readout electronics for the 80~L Rn detector are similar to those used in earlier detectors~\cite{Radon_takeuchi_2003}.
However, the previous charge sensitive preamplifier, CLEAR PULSE CS513-1, has been replaced with the CS515-1 model.
 Working with CLEAR-PULSE Co. Ltd,  a new data acquisition system has been developed, which 
not only records signals from the Rn detector, but also monitors several environmental parameters which affect the Rn detection efficiency, 
such as the temperature, dew point, and flow rate of the sample air.
The data logger (CP 80278) consists of one CPU board (CP 1233), two shaping amplifier boards (CP 4485), two pulse height analyzer boards (CP 1234), and two analog-to-digital converter (ADC) boards (CP 1235).

\subsection{Radon spectrum and event selection}
\label{sec:rnspec}
\begin{figure}[h]
 \begin{center}
  \includegraphics[width=120mm]{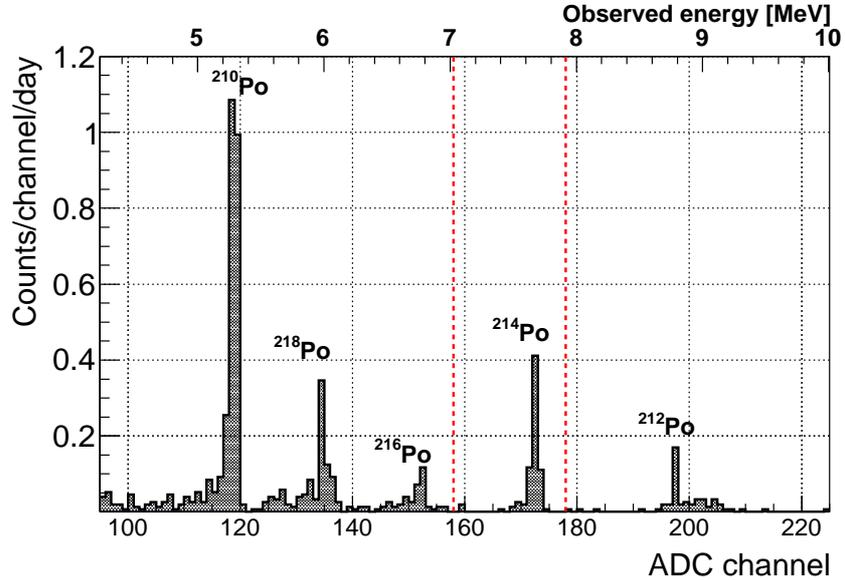}
 \end{center}
 \caption{Typical energy spectrum from $^{222}\mathrm{Rn}$ and $\mathrm{^{220}Rn}$ daughters. The region between ADC channels 158 to 178 and shown by the red dashed lines defines the $^{214}\mathrm{Po}$ signal region used to calculate the $\mathrm{^{222}Rn}$ concentrations for this particular detector.}
 \label{spec}
\end{figure}

An integrated pulse height spectrum from a background run performed with pure air (c.f. Sec~\ref{sec4}) is shown in Fig. \ref{spec}. 
There are five peaks in the spectrum originating from either $^{222}$Rn or $^{220}$Rn daughters. Here, $^{210}$Po, $^{214}$Po and $^{218}$Po are the decay products of $^{222}$Rn while $^{212}$Po and $^{216}$Po are the decay products of $^{220}$Rn. In the measurement presented below the $^{214}$Po signal ($7.69$ $\mathrm{MeV}$) is used to calculate $\mathrm{^{222}Rn}$ concentrations because there are no other $\alpha$ sources with energies near its peak value; the closest being the 6.78~MeV $\alpha$ from $\mathrm{^{216}Po}$. In addition, $^{214}$Po is known to have a higher collection efficiency than $^{218}$Po~\cite{Radon_takeuchi_1999,Radon_takeuchi_2003,PTEP_hoso}.

The $\mathrm{^{214}Po}$ signal region in the pulse height spectrum is determined using the following criteria. In order to prevent contamination by $^{216}\mathrm{Po}$ events due to the resolution of the PIN photodiode, a lower bound on the signal region is defined as five ADC channel above the peak position of the $^{216}\mathrm{Po}$ signal. Loss of higher energy $^{214}\mathrm{Po}$ signals is prevented using an upper bound, chosen to be five ADC channels above the $^{214}\mathrm{Po}$ peak position.
These cuts provide near 100\% selection efficiency for the signal with minimal backgrounds from the tail of the $^{216}\mathrm{Po}$ spectrum. However, we revised the selection criterion in order to accumulate more $\mathrm{^{214}Po}$ signals. Contamination from the latter represents a negligible source of measurement uncertainty in comparison to other systematic errors discussed below. We note that in the previous publication \cite{PTEP_hoso}, the range of ADC channels defining the signal region was determined using other criteria and resulted in a $\mathrm{^{214}Po}$ detection efficiency of only $90\%$. For the spectrum shown in Fig.~\ref{spec}, the signal region for $^{214}$Po $\alpha$ decay is taken to be ADC channels 158 to 178, which corresponds to energies between $7.03$ and  $7.92$ MeV.
 
\section{Detector calibration}
\label{sec3}
\subsection{Calibration factor}
In order to calculate the $\mathrm{^{222}Rn}$ concentration from the observed $\mathrm{^{214}Po}$ count rate, the 80~L Rn detector must be calibrated.  
A calibration factor ($C_{F}$) is defined as follows:
\begin{equation}
C_{F}\mathrm{[(counts/day)/(mBq/m^{3})]  = \frac{measured\mspace{5mu} {}^{214}Po \mspace{5mu}signal\mspace{5mu} rate}{^{222}Rn \mspace{5mu}concentration}}. \label{cf}
\end{equation}
\noindent Here, the numerator is the measured $^{214}$Po signal rate on the PIN photodiode in units of $\mathrm{counts/day}$, while the denominator is the $^{222}$Rn concentration to be determined in the unit of $\mathrm{mBq/m^{3}}$. 
The $C_{F}$ forms the core of the following measurements, and its initial value is calibrated using a Rn source of well-measured activity as described below.

\subsection{Calibration setup for 80 L Rn detector}
\begin{figure}[h]
\centering
\includegraphics[width = 120mm]{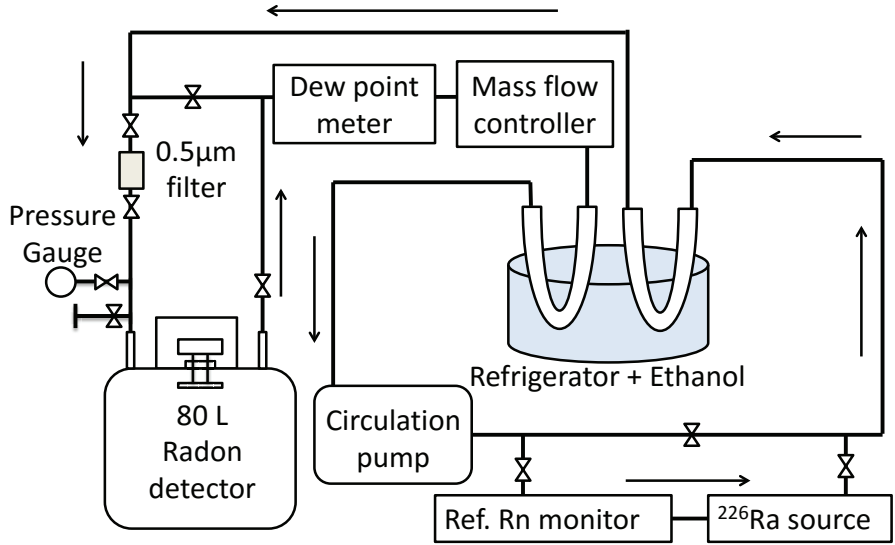}
\caption{Schematic diagram of the calibration system. Arrows indicate the direction of air flow.}
\label{calib}
\end{figure}

The calibration system constructed at Kamioka observatory is shown in Fig.~\ref{calib}.
 Using this calibration system, both the high-voltage and humidity dependence of the $C_{F}$ were measured (see also~\cite{PTEP_hoso}).

The system consists of the 80~L Rn detector, a cold bath to control the dew point in the target air, 
an air circulation pump, a dew-point meter (VAISALA, DMT340), a mass flow controller (HORIBA STEC Z512), 
a reference radon monitor (ionization chamber, SAPHYMO Alpha GUARD PQ2000), 
and a $^{222}$Rn source (PYLON RNC).
During calibrations $^{226}$Ra with an activity of 199 Bq $\pm 4\%$ is used to maintain radioactive equilibrium within the entire system.
Two stainless 0.5~inch U-shaped pipings are used as heat exchangers to control the system's dew-point.
They are placed in an ethanol bath with temperature control provided by a refrigerator (THOMAS,TRL-107SLC).
 The temperature and dew point of the air through the entire system are monitored by the dew point meter.
To suppress emanation of $\mathrm{^{222}Rn}$ from the piping,  all piping used in the system is made of electropolished stainless steel (SUS EP tube). 
All connections use VCR$^{\textcircled{\footnotesize{R}}} $ gaskets to prevent air leakage and contamination from environmental air.
Commercially-available G1-grade high-purity air (impurity concentration  $<0.1$ ppm) is used to minimize background $\mathrm{^{222}Rn}$ in the calibration fill air itself. 
The fill air is continuously circulated at 1 atm with a flow rate of $0.8$~L/min controlled by the mass flow controller. It should be noted that the system's pump is only capable of circulating air at this flow rate.

\subsection{High-voltage dependence of the calibration factor}
\begin{figure}[t]
 \begin{center}
  \includegraphics[width=120mm]{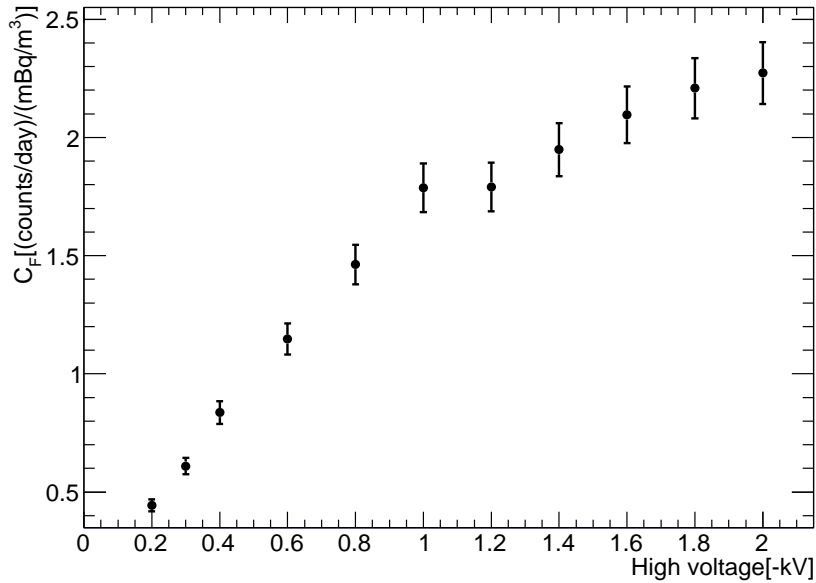}
 \end{center}
 \caption{ High-voltage dependence of the $C_{F}$ for the electrostatic collection method of pure air.
 These data were obtained at an absolute humidity of $0.0021$ $\mathrm{g/m^{3}}$.
 The bars show both statistical and systematic errors.}
 \label{hv_dep}
\end{figure}

As an electric field is formed between the PIN photodiode and the stainless steel vessel by the supplied high voltage, the electrostatic collection efficiency directly depends on the supplied high voltage.
 Accordingly, the detector calibration was conducted with various  supply voltages to test this dependence.
 First, the entire system was evacuated down to $1.0\times10^{-5}$~Pa without the reference Rn monitor and without the $^{226}$Ra source.
 Then, the system was filled with G1-grade air at atmospheric pressure and the valves in front of and behind the $^{226}$Ra source were opened and air circulation was started.
 Finally, after the closed system reached radioactive equilibrium, the Rn concentration measured by the reference Rn monitor and the number of $^{214}$Po counts in in the 80~L Rn detector were compared in order to determine the $C_{F}$.
 The supplied high voltage was varied from $-0.2$ to $-2.0$ kV during this calibration.
 Fig.~\ref{hv_dep} shows the applied high-voltage dependence of the $C_{F}$.
 During calibration, the dew point was maintained at a value measured to be $0.0021$ $\mathrm{g/m^{3}}$.

Although the $C_{F}$ is not saturated at $-2.0$~kV, this voltage was chosen for other calibrations and measurements because it is the highest that can be applied.
However, the circuit's high-voltage divider also supplies a reverse bias to the PIN photodiode~\cite{Radon_takeuchi_2003}, 
which increases as the supplied high voltage increases.
The bias voltage is $-0.1$~kV, the limit of the photodiode's tolerance, at a supply voltage of $-2.0$~kV. 
This configuration therefore provides $-1.9$~kV for the generation of the detector's electrostatic field.

\subsection{Humidity dependence of the calibration factor}
\begin{figure}[h]
 \begin{center}
  \includegraphics[width=120mm]{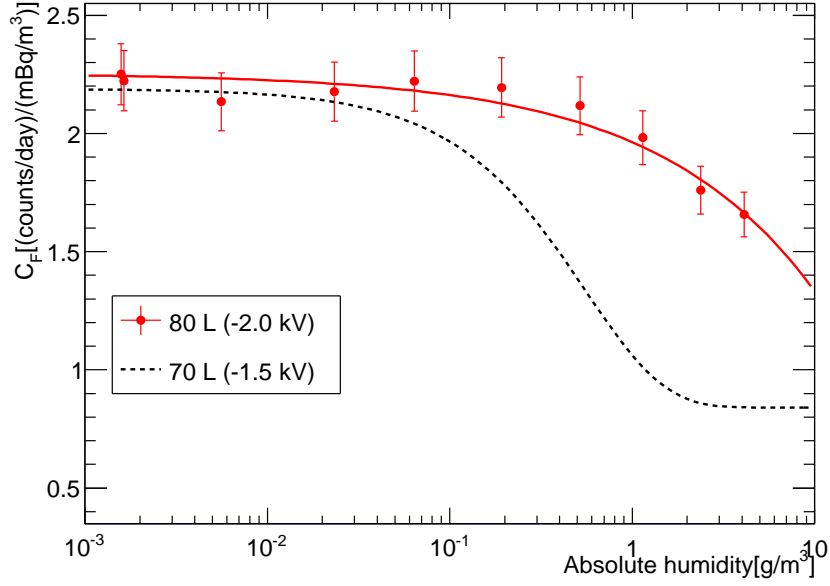}
 \end{center}
 \caption{Absolute humidity dependence of the $C_{F}$ (red).
 Calibration data for the 80 L~Rn detector were taken at $-2.0$ kV.
 The bars show both statistical and systematic errors. The solid line (red) shows the best-fit function.
 The dashed line (black) shows the old result, which was obtained by the old 70 L Rn detector at $-1.5$ kV \cite{Radon_takeuchi_1999}.}
 \label{ah_dep}
\end{figure}

Because the positively charged $^{222}$Rn daughter nuclei are captured and neutralized by water in air, the detection efficiency of the Rn detector also depends on the absolute humidity~\cite{neut}.
 Therefore, calibrations to determine the humidity dependence of the $C_{F}$ were conducted using the same setup and settings used for the high-voltage dependence calibration.
 After the system reached radioactive equilibrium, the dew-point temperature was controlled by the refrigerator in order to vary the absolute humidity throughout the entire system. 
Fig.~\ref{ah_dep} shows the absolute humidity dependence of the 80~L Rn detector's $C_{F}$ for air, which is seen to decrease at higher absolute humidities as expected.
 While the $C_{F}$ for earlier 70~L detectors was found to be constant above 1.5 $\mathrm{g/m^{3}}$, the current detector demonstrates a humidity dependence even above this value. Relative to the previous design, the current detector's larger $C_{F}$ at absolute humidities above $10^{-1}$ $\mathrm{g/m^{3}}$ is a consequence of the higher vacuum its CF flanges provide.  The data are found to be best described by an empirical function~\cite{nemoto} $C_{F} = 2.25-0.29\sqrt{A_{H}}$, where $A_{H}$ is the absolute humidity.

\subsection{Uncertainty budget for the detector calibration}
 Several sources of systematic uncertainty are present in the calibration of the $C_{F}$. The reference monitor has an uncertainty of  $\pm5\%$. According to the manufacturer's documentation, the uncertainty of the Alpha GUARD calibration is $\pm3\%$ \cite{alpha_guard}. 
However, in order to confirm this uncertainty, we compared the expected Rn concentration of the measurement system in radioactive equilibrium with 
a calibration source with that measured by the Alpha GUARD monitor and found them to agree within $\pm5\%$.
As a result this value is used for the uncertainty of the reference monitor in the present analysis.
 In addition the dew-point meter has an uncertainty of $\pm2\%$ and the uncertainty 
 arising from the estimation of the total volume in the calibration setup due to, for instance, the uncertainty in the volume of its flexible tubing, is taken to be $\pm2\%$. Based on these uncertainties, the total systematic uncertainty is taken to be $\pm5.7\%$ for each $C_{F}$ measurement. The uncertainty for the calibration is summarized in Table \ref{budget}.
\begin{table}[]
  \begin{center}
    \caption{Summary of systematic uncertainties}
    \begin{tabular}{|c|c|c|c|} \hline
      Source & Measured value & Assigned standard uncertainty\\ \hline \hline
	Reference Rn monitor  & 2.2 $\mathrm{kBq/m^{3}}$ & 0.1 $\mathrm{kBq/m^{3}}$ \\
	Dew point meter & $-75$ $\mathrm{^{\circ}C}$ & $\pm(2{\sim}3)$ $\mathrm{^{\circ}C}$ \\
	Estimation of the whole volume  & 85~L & 2~L   \\ \hline \hline
     Subtracting background &  $0.33$ $\mathrm{mBq/m^{3}}$ & $0.07$ $\mathrm{mBq/m^{3}}$ \\ \hline
    \end{tabular}
    \label{budget}
  \end{center}
\end{table}

\section{Intrinsic background levels}
\label{sec4}
After these calibrations, the background level intrinsic to the 80~L Rn detector itself was evaluated.
Data were taken over 156 days with no input Rn source. 
During this time period the average absolute humidity was $0.078$ $\mathrm{g/m^{3}}$ and the observed $0.74 \pm 0.07$ (statistical uncertainty only) counts/day for the $^{214}\mathrm{Po}$ decay corresponds to 
\begin{equation}
C_{\mathrm{detector,BG}} = 0.33\pm0.07 \mspace{5mu}\mathrm{mBq/m^{3}}.
\label{bg_eq}
\end{equation} 
This intrinsic background level is lower than that of the previous 70~L detectors ($2.4\pm1.3$ count/day)~\cite{Radon_takeuchi_1999} by a factor of three. The characteristic detection limit of the detector is estimated as $L_{c}=1.64\times\sigma_{\mathrm{BG}}/C_{F}$~[$\mathrm{mBq/m^{3}}$], where $\sigma_{\mathrm{BG}}$ is the uncertainty of the background rate~\cite{Radon_takeuchi_2003}, which results in $0.54$~$\mathrm{mBq/m^{3}}$ for a single day measurement. The energy spectrum from the measurement is shown in Fig. \ref{spec} and is discussed further in later Sections. 

\section{Measurement of the Rn concentration in the Super-K buffer gas layer}
\label{sec5}
\subsection{Introduction}
In the Super-K tank, there is a buffer gas layer between the surface of the water and the top of the tank~\cite{paper_SK} to prevent the tank seal from breaking due to the motion of its water during an earthquake.
 The height of the buffer gas layer is 60~cm and its total volume is $753.6$ $\mathrm{m^{3}}$. 

Rn-reduced air system~\cite{Radon_takeuchi_2003} 
supplies purified air for the Super-K buffer gas layer, injecting it into the tank from an inlet at $(x,y)=(+18.382$ $\mathrm{m},$ $-4.242$ $\mathrm{m})$ in the detector coordinates.
This system provides a gas flow rate of $11$ $\mathrm{m^{3}/h}$ and maintains the pressure of the buffer gas layer at $+0.3$ kPa relative to atmospheric pressure. Since the top of the Super-K tank is exposed to the dome of its experimental hall, where the Rn concentration is $40$--$100$ $\mathrm{Bq/m^{3}}$, continuous purging of the buffer gas in this manner is necessary to suppress Rn emanation into the buffer layer.

\subsection{Rn-reduced air system improvements}
The Super-K Rn-reduced air system uses 50~L of chilled activated charcoal to remove Rn from throughput air.
Since it is known that the Rn-trapping efficiency of activated charcoal increases significantly below temperatures around $-60$ $\mathrm{^{\circ}C}$~\cite{charcoal}, 
the system was upgraded in March 2013 to bring its cooling power down from $-40$ $\mathrm{^{\circ}C}$~\cite{Radon_takeuchi_2003} in the previous design, to below this threshold.
In order to do so, the system's coolant was changed to 3M $\mathrm{NOVEC^{TM}}$7100 and its refrigerator upgraded. 

\subsection{The monitoring system}
\begin{figure*}[]
\begin{center}
\includegraphics[width=100mm]{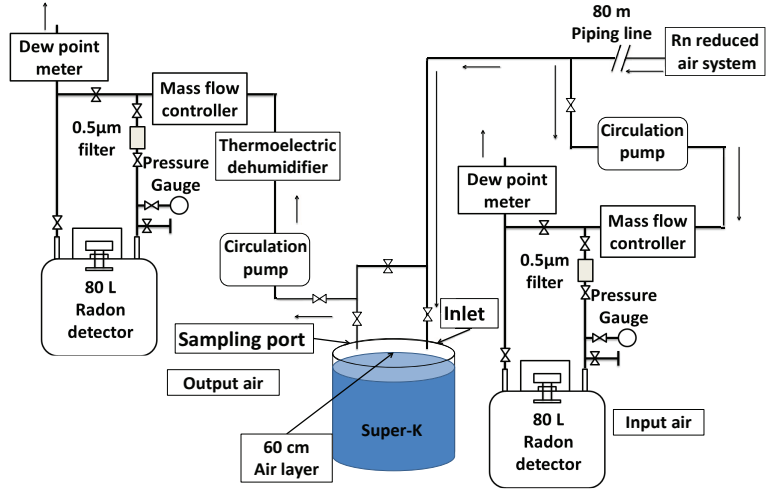}
\end{center}
\caption{Schematic of the Rn concentration monitoring system for the Super-K buffer gas.}
\label{air_system}
\end{figure*}

An upgraded measurement system used for the measurements of the buffer gas Rn concentration is shown in Fig.~\ref{air_system}.
 Using this system, the Rn concentration both before and after the activated charcoal cooling system upgrade was measured.
Two 80~L Rn detectors were set on the top of the Super-K tank, one to measure the Rn concentration of the gas supplied to the tank (input air), and the other to measure gas sampled from the buffer layer (output air).
 The output air was sampled from the sampling port at $(x,y) = (-14.494$ $\mathrm{m},$ $+12.019$ $\mathrm{m})$, which is $\sim$ 23 m away from the buffer gas inlet.
 Both the input and output air were sampled using circulation pumps with flow rates set by the mass flow controllers to  $0.8$ $\mathrm{L/min}$.

Because the absolute humidity of the output air is above 5 $\mathrm{g/m^{3}}$, a point not measured during the $C_{F}$ calibration, 
a thermoelectric dehumidifier (KELK DH-209C) was installed just before the Rn detector to reduce the humidity, as shown in Fig. \ref{air_system}. The absolute humidity after the installing the dehumidifier was $\sim3.5$ $\mathrm{g/m^{3}}$.

\subsection{Results}
\begin{figure}[h]
 \begin{center}
  \includegraphics[width=120mm]{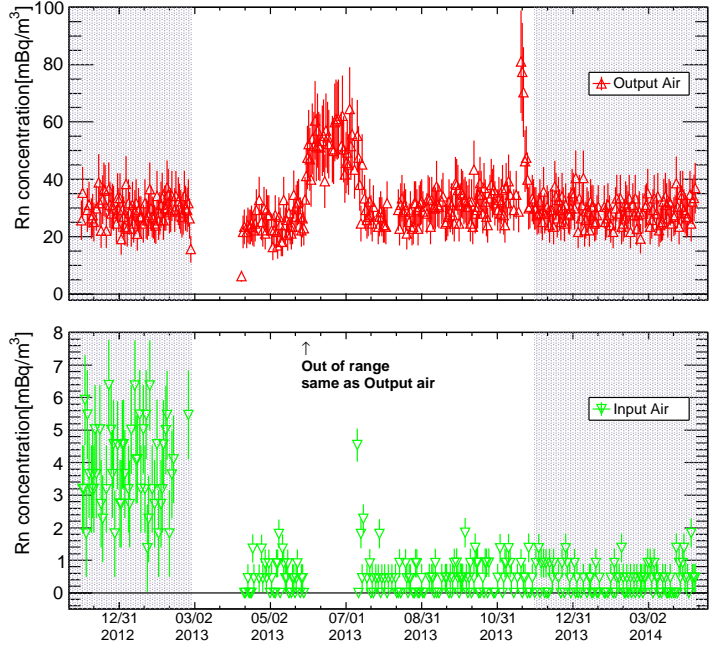}
 \end{center}
 \caption{Rn concentration in the input air (light-green) and output air (red) of the Super-K detector's buffer layer.
 The data set is from December $\mathrm{1^{st}}$, 2012 to April $\mathrm{9^{th}}$, 2014.
 Data from the shaded regions were used for the analysis of concentrations before and after the improvement of the Rn-reduced air system. Error bars show both statistical and systematic errors.}
 \label{air_result_before}
\end{figure}

\begin{figure}[h]
 \begin{center}
  \includegraphics[width=120mm]{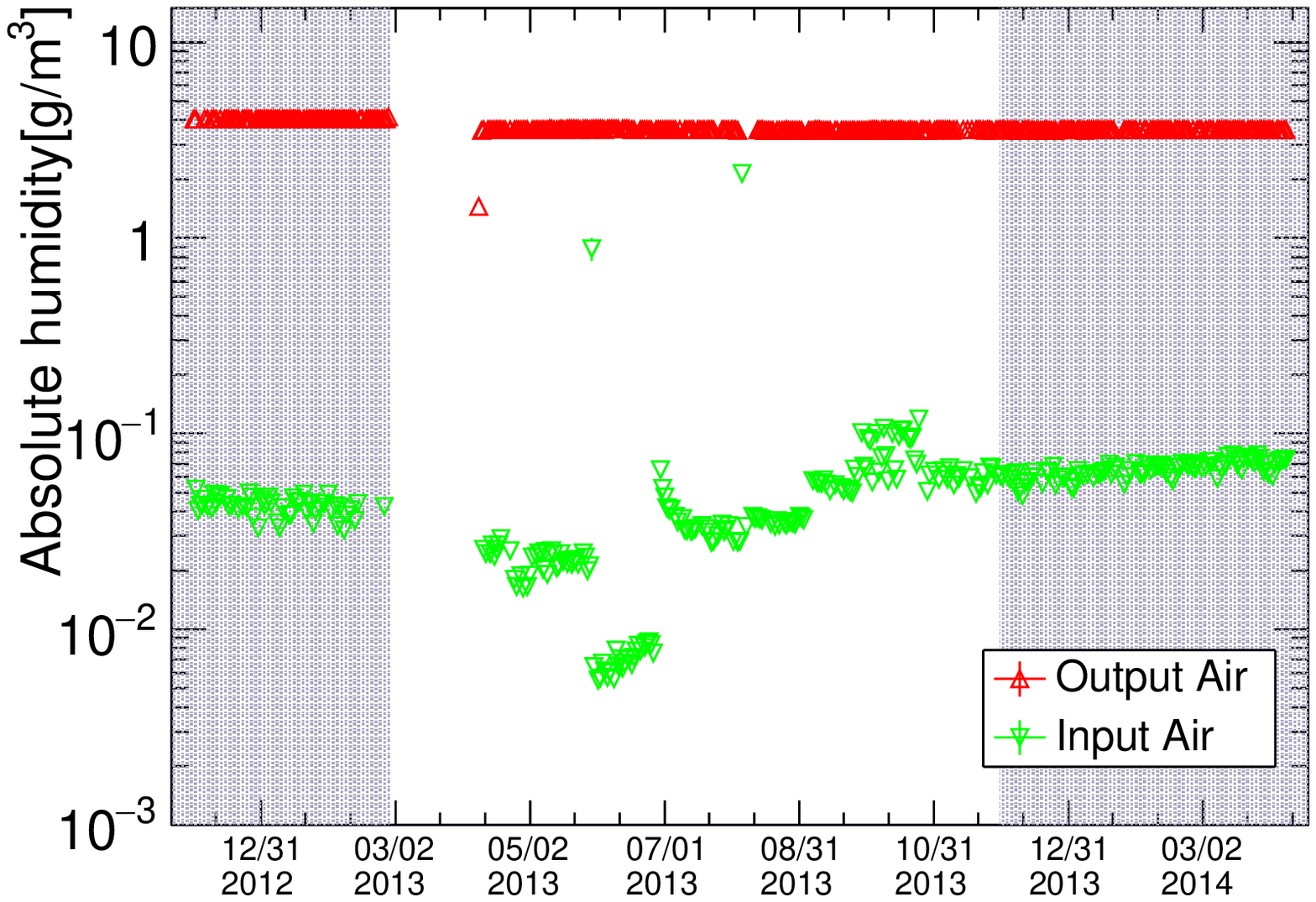}
 \end{center}
 \caption{Absolute humidity from the input (light-green) and output (red) December $\mathrm{1^{st}}$, 2012 to April $\mathrm{9^{th}}$, 2014.
 Data from the shaded regions were used for the analysis of concentrations before and after the improvement of the Rn-reduced air system.}
 \label{humidity_result_bef}
\end{figure}

Monitoring of the Rn concentration in the buffer gas began on December $\mathrm{1^{st}}$, 2012.
 The results are summarized in Fig.~\ref{air_result_before} with the absolute humidity of the sampled air during the same period shown in Fig.~\ref{humidity_result_bef}.
 Gaps in the data are due to intermittent problems, including a laboratory-wide power failure, network connection issues, and problems with the data acquisition system.
During the period from May $\mathrm{31^{st}}$ to July $\mathrm{11^{th}}$, 2013 
failure of an automatic valve in the Rn-reduced air system resulted in the buffer gas accidentally bypassing the chilled activated charcoal system.
 This in turn caused the Rn concentrations in both the input and the output air to increase during this period.
On September $\mathrm{16^{th}}$, 2013, a power failure occurred as a result of a thunder storm, and the activated charcoal was not subject to cooling for more than four hours.
 After we resumed running the system, the absolute humidity of the input air became higher than before but, fortunately, the Rn concentration did not change.
In addition, on November $\mathrm{9^{th}}$, 2013, a scheduled laboratory-wide power outage cut power to the Rn-reduced air circulation system.  When the system was restarted, unpurified air was accidentally supplied to the buffer layer, causing the Rn concentration in the output air to increase for a few days.

Over the 90 days prior to the improvement of the Rn-reduced air system the input and output air Rn concentrations were measured to be,
\begin{align*}
C_{\mathrm{Input,before}} &= 3.5\pm0.3 \mspace{5mu}\mathrm{mBq/m^{3}}
\;\;\;\mbox{and} \\ 
C_{\mathrm{Output,before}} &= 28.9\pm1.7\mspace{5mu}\mathrm{mBq/m^{3}},
\end{align*}
\noindent respectively.
These values were derived after subtracting $C_{\mathrm{detector,BG}}$.
It should be noted that the Rn concentration in the output air was found to be much higher than that of the input air both before and after the improvement, which confirms that the buffer gas layer is contaminated with Rn. That is, a Rn source exists inside the Super-K tank. Although the origin of this source has not yet been identified, the most likely candidate is emanation from the detector structure.

\begin{figure}[h]
 \begin{center}
  \includegraphics[width=120mm]{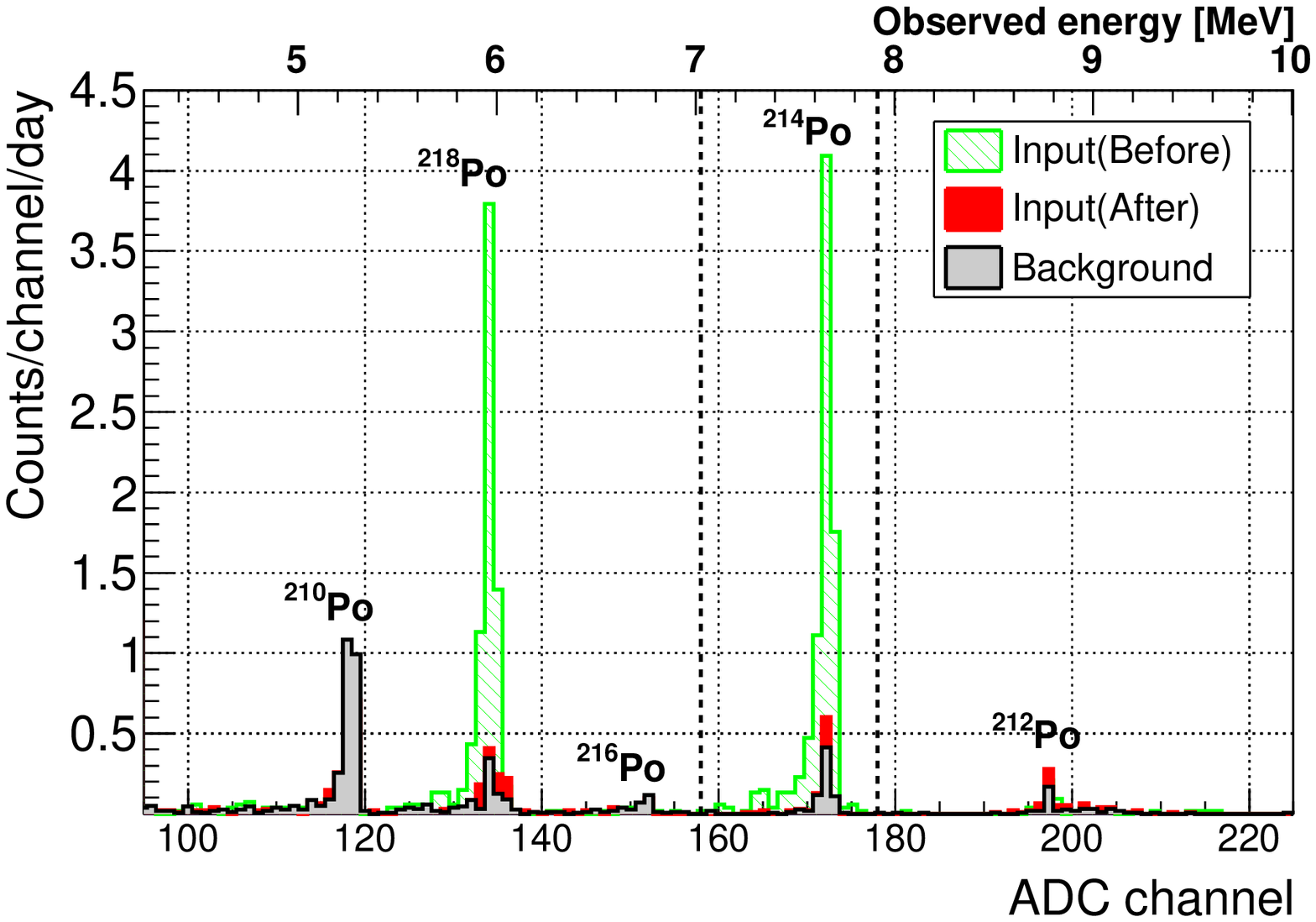}
 \end{center}
 \caption{Spectrum measurements using a single 80~L Rn detector. 
The gray spectrum corresponds to the background measurement, while the light-green (red) corresponds to the input air measurement before (after) the air system improvement. 
The $\mathrm{^{214}Po}$ signal region for this detector is defined as counts between ADC channels 158 and 178, corresponding to energies between 7.03 and 7.92 MeV, 
based on the procedure described in Sec~\ref{sec:rnspec}. 
}
 \label{spec_bg}
\end{figure}

Data from December $\mathrm{1^{st}}$, 2013 to April $\mathrm{9^{th}}$, 2014 were used to evaluate the Rn concentrations after the improvement of the Rn-reduced air system, 
During this period the operation of the Rn-reduced air system and the data acquisition were stable.
The results for measurements taken 131 days after the system improvement are, 
\begin{align*}
C_{\mathrm{Input,after}} &= 0.08 \pm0.07 \mspace{5mu} \mathrm{mBq/m^{3}}
\;\;\;\mbox{and} \\ 
C_{\mathrm{Output,after}} &= 28.8\pm1.7\mspace{5mu}\mathrm{mBq/m^{3}},
\end{align*}
\noindent after background subtraction.
Energy spectra from measurements of the input air are shown in Fig.~\ref{spec_bg}.
It can be seen that the Rn daughters in the supplied buffer gas have been reduced by a factor of four following the air system upgrade.
Therefore, it can be concluded that the new $-60$ $\mathrm{^{\circ}C}$ chilled activated charcoal works efficiently to reduce Rn and the product gas has become effectively free of Rn.
On the other hand, the Rn concentration of the output air after the improvement has not changed.
This indicates that contamination from the Super-K tank is the dominant source of Rn in the buffer gas.

It should be noted that the count rate of $\mathrm{^{210}Po}$ in the background spectrum is larger than the other measurements because it was performed last. The increase is likely due to positively charged $\mathrm{^{210}Pb}$ from the $\mathrm{^{222}Rn}$ decay chain accumulating on the surface of the PIN photodiode over the measurement history of the detector. Since it decays with a 22-year half-life to produce the observed $\mathrm{^{210}Po}$, the increased count rate thus indicates that $\mathrm{^{222}Rn}$ daughters remain on the photodiode once they have been collected.

\section{Conclusion and future prospects}
\label{sec6}
In order to measure the Rn concentration in the buffer gas used in Super-Kamiokande, a new 80~L Rn detector has been developed and calibrated onsite at Kamioka observatory using pure air. Typical calibration factors for the detector are $2.25\pm0.13$ $\mathrm{CPD/(mBq/m^{3})}$ 
at $0.002$ $\mathrm{g/m^{3}}$ and $1.66\pm0.14$ $\mathrm{CPD/(mBq/m^{3})}$ at $4.1$ $\mathrm{g/m^{3}}$.
Compared to previous designs the current detector has an extremely low intrinsic background, measured to be $0.33\pm0.07$ $\mathrm{mBq/m^{3}}$.
Using this detector, the Rn concentration of the buffer gas in the Super-K tank has been measured precisely.
Following improvements to the Rn-reduced air system, the Rn concentration of the supplied buffer gas was reduced to $0.08\pm0.07$ $\mathrm{mBq/m^{3}}$.
However, the concentration in the buffer gas layer itself has been measured to be $28.8\pm1.7$ $\mathrm{mBq/m^{3}}$ suggesting 
the existence of Rn sources in the top region of the Super-K tank.

As a next step, the Rn concentration in the Super-K purified water is being measured in order to evaluate the 
contribution of background events originating from Rn daughters to solar neutrino analysis.
For this purpose, a system based on the new 80~L detector and capable of measuring Rn concentrations in purified water 
at the level of $\sim0.1$ $\mathrm{mBq/m^{3}}$ has been developed~\cite{nakano}.
Details of the system design and measurements of the Rn concentration in the Super-K tank water will be reported in a subsequent publication.
Several other applications using the 80~L Rn detector are ongoing, including Rn concentration measurements in several novel gases~\cite{PTEP_hoso} and Rn emanation measurements from solid materials for detector material screening~\cite{sekiya}.

\section*{Acknowledgments}
The authors would like to thank the Super-Kamiokande collaboration for their help in conducting this study. We gratefully acknowledge the cooperation of the Kamioka Mining and Smelting Company. Y. N thanks M. Kanazawa for assembling both the calibration and the measurement setups. This work is partially supported by MEXT KAKENHI Grant Number JP26104008 and the inter-university research program at ICRR.





\end{document}